# "Four Ways to TeV Scale", Ankara'97 Workshop and Something Else

S. Sultansoy [a, b, c]

[a)] *Deutsches Elektronen-Synchrotron DESY, Notke Str. 85, D-22607 Hamburg, GERMANY*

[b)] *Physics Dept., Faculty of Arts and Sciences, Gazi Univ., 06500 Ankara, TURKEY*

[c)] *Institute of Physics, Academy of Sciences, H. Cavid Avenue 33, Baku, AZERBAIJAN*

**Abstract**

The first Part contains the invited talk titled "Four Ways to TeV Scale" given at the First International Workshop on Linac-Ring Type ep and $\gamma$p Colliders (9-11 April 1997, Ankara, Turkey). Four known types of colliders, which may give an opportunity to achieve TeV center of mass energies in the near future (10-15 years), are discussed. Parameters of the linac-ring type *ep, $\gamma$p, e-nucleus and $\gamma$-nucleus* machines are roughly estimated. Some speculations on TeV scale physics are given. The physics goals of the TeV energy $\gamma$p and $\gamma$-nucleus colliders are considered.

Information about the Ankara'97 Workshop, including Workshop Conclusion and Recommendations, is presented in Part II. Finally, an Appeal to ICFA, ECFA and ACFA is presented in Part III.

**CONTENTS**

**PREFACE**

# PREFACE

It is well known that lepton-hadron collisions play a crucial role in our understanding of micro-world. An excellent example is the discovery of the quark-parton structure of nucleons. The fixed target experiments explore the region $Q^2 < 1000$ GeV$^2$. The HERA collider enlarged this region by two orders. The possible LEP*LHC collider has an important disadvantage of $E_e/E_p < 0.01$. The synchrotron radiation restricts the electron energy available at ring machines and a transition to linear accelerators seems unavoidable for $E_e > 100$ GeV. Therefore, one should consider linac-ring type machines in order to achieve TeV scale at constituent level in lepton-hadron collisions (see Part I of this paper). A possible alternative, namely µp collider will face even more problems than the basic ($\mu^+\mu^-$) collider.

Linac-ring type colliders present the fourth way to achieve TeV center of mass energies at constituent level in the near future (10-15 years). However, an intensive work on both the machine parameters and the physics search potential is needed. As distinct from other three ways (pp, $e^+e^-$ and $\mu^+\mu^-$ colliders), linac-ring type machines did not attract adequate attention of the world high energy physics community, as well as corresponding organizations. The first (and last) International meeting on the subject held in Ankara three years ago. Let me cite the last sentence of Ankara'97 Workshop's Conclusion and Recommendations (see Part II of this paper):

*As a result of the workshop, participants came to the point that it will be useful to organize two workshops, one on the machine parameters and the other on the physics research program, during the next year.*

Unfortunately, these recommendations were not realized.

This paper presents my last attempt to draw an attention of the ICFA and so on in order to initiate more activity on the subject (see Part III).

# PART I. "Four Ways to TeV Scale"

# Four Ways to TeV Scale[1]

S. Sultansoy

*Department of Physics, Faculty of Sciences, Ankara University, 06100 Tandogan, Ankara, Turkey*
and
*Institute of Physics, Academy of Sciences, H. Cavid avenue 33, Baku, Azerbaijan*

**Abstract**

Four known types of colliders, which may give an opportunity to achieve TeV center of mass energies in the near future (10-15 years), are discussed. Parameters of the linac-ring type $ep$ and $\gamma p$ machines are roughly estimated. Some speculations on TeV scale physics are given. The physics goals of the TeV energy $ep$ and $\gamma p$ colliders are considered.

---

[1] Talk given at the First International Workshop on Linac-Ring Type $ep$ and $\gamma p$ Colliders (9-11 April 1997, Ankara, Turkey). Published in Turkish J. of Phys. 22 (1998) 575-594.

# 1. Introduction

It is known that the Standard Model with three fermion families well describes almost the entire large amount of particle physics phenomena [1]. Today, SM is proved at the level of first-order radiative corrections for energies up to 100 GeV. However, there are a number of fundamental problems which do not have solutions in the framework of the SM: quark-lepton symmetry and fermion's mass and mixings pattern, family replication and number of families, L-R symmetry breaking, electroweak scale etc. Then, SM contains unacceptably large number of arbitrary parameters even in three family case: 19 in the absence of right neutrinos (and Majorana mass terms for left neutrinos), 26 if neutrinos are Dirac particles, $\geq 30$ if neutrinos are Majorana particles. Moreover, the number of "elementary particles", which is equal to 37 in three family case (18 quarks, 6 leptons, 12 gauge bosons and 1 Higgs boson), reminds the Mendeleev Table. Three decades ago similar situation led to the quark model!

For these reasons, physicists propose a lot of different extensions of the SM, most part of which predicts a rich spectrum of new particles and/or interactions at TeV scale. These extensions can be grouped in two classes, namely standard and radical ones. Standard extensions remain in the frameworks of gauge theories with spontaneously broken gauge symmetry and include an enlargement of Higgs sector, enrichment of fermion sector, introducing of new gauge symmetries etc. Radical extensions include: compositness (preonic→pre-preonic models?), SUSY (MSSM→SUGRA) and "unexpected" new physics (new space-time dimensions at TeV scale etc.).

An exploration of TeV region will require all possible types of colliding beams in order to clarify true new physics.

# 2. TeV Energy Colliders

Today, there are four (more or less known) types of colliders, which may give opportunity to achieve TeV center of mass energies at constituent level in the near future (10-15 years):

i) Hadron colliders, namely LHC [1] (and may be Upgraded FNAL)
ii) Linear $e^+e^-$ colliders [2] (including $\gamma e$ and $\gamma\gamma$ options)
iii) Linac-ring type ep ($\gamma p$) colliders
iv) $\mu^+\mu^-$ colliders [3].

The first two are well known. The third type is less known: this workshop is the first international one held on the subject. The fourth type is sufficiently well known, because a number of workshops and conferences on this subject were held during last years.

Physics search programs of these machines are complementary to each other and construction of all of them will give opportunity to investigate TeV scale in the best manner.

## 2.1. The (center of mass) Energy Frontiers

Today we have the following situation:

### 2.1.1. Hadron Colliders
TEVATRON (Fermilab) $\bar{p}p$
$\sqrt{s} = 2$ TeV $\to$ 4 TeV ??
$L = 2.5 \cdot 10^{31}$ cm$^{-2}$s$^{-1}$ $\to 10^{33}$ cm$^{-2}$s$^{-1}$ (2000)

### 2.1.2. Lepton Colliders
LEP (CERN) $e^+e^-$: $\sqrt{s} = 180$ GeV and $L = 2.4 \cdot 10^{31}$ cm$^{-2}$s$^{-1}$
SLC (SLAC) $e^+e^-$: $\sqrt{s} = 90$ GeV and $L = 0.8 \cdot 10^{31}$ cm$^{-2}$s$^{-1}$

### 2.1.3. Lepton-Hadron Colliders
HERA (DESY) $e^{\pm}p \to e^{\pm}$–nucleus ?
$\sqrt{s} = 300$ GeV
$L = 1.6 \cdot 10^{31}$ cm$^{-2}$s$^{-1}$ $\to 10^{32}$ cm$^{-2}$s$^{-1}$ (1998)

We hope that during the next decade following machines will be constructed:

Hadron colliders: LHC (CERN) pp with $\sqrt{s} = 14$ TeV and $L = 10^{34}$ cm$^{-2}$s$^{-1}$.

Lepton colliders:
NLC (DESY, KEK) $e^+e^-$ ($\gamma e$, $\gamma\gamma$) with $\sqrt{s} = 0.5$ ($\to 1.5$) TeV and $L = 10^{34}$ cm$^{-2}$s$^{-1}$.
$\mu^+\mu^-$ (USA) with $\sqrt{s} = 0.5$ TeV and $L = 10^{33}$ cm$^{-2}$s$^{-1}$.

Lepton-hadron colliders: ??

## 2.2. Accelerators for Physics Studies

### 2.2.1. Snowmass '96
Following colliders have been proposed in order to discuss their physics search goals:

- Tevatron
  - $E_{cm} = 2$ TeV, $L = 10^{33}$ cm$^{-2}$s$^{-1}$
- LHC
  - $E_{cm} = 14$ TeV, $L = 10 \cdot 10^{33}$ cm$^{-2}$s$^{-1}$
- NLC
  - $E_{cm} = 0.5$ TeV, $L = 5 \cdot 10^{33}$ cm$^{-2}$s$^{-1}$
  - $E_{cm} = 1$ TeV, $L = 20 \cdot 10^{33}$ cm$^{-2}$s$^{-1}$
  - $E_{cm} = 1.5$ TeV, $L = 20 \cdot 10^{33}$ cm$^{-2}$s$^{-1}$
- $\mu^+\mu^-$
  - $E_{cm} = 0.5$ TeV, $L = 0.7 \cdot 10^{33}$ cm$^{-2}$s$^{-1}$
  - $E_{cm} = 0.5$ TeV, $L = 5 \cdot 10^{33}$ cm$^{-2}$s$^{-1}$
  - $E_{cm} = 4$ TeV, $L = 100 \cdot 10^{33}$ cm$^{-2}$s$^{-1}$

- pp, VLHC
  - $E_{cm}$ = 60 TeV, L = 10·10$^{33}$cm$^{-2}$s$^{-1}$
- e$^+$e$^-$, LSC (Linear Super Collider)
  - $E_{cm}$ = 5 TeV, L = 100·10$^{33}$cm$^{-2}$s$^{-1}$
- ep, LHC×LEP ?
  - $E_{cm}$ = 1 TeV, L = 0.1·10$^{33}$cm$^{-2}$s$^{-1}$

### 2.2.2. Linac-Ring Type *ep* and γ*p* Colliders

In our opinion following ep and γp colliders should be added to this list:

- HERA⊗NLC
  - $E_{cm}$ = 1 TeV, L = 0.1·10$^{33}$cm$^{-2}$s$^{-1}$
  - $E_{cm}$ = 2.4 TeV, L = 0.1·10$^{33}$cm$^{-2}$s$^{-1}$
- LHC⊗NLC
  - $E_{cm}$ = 2.6 TeV, L = 0.5·10$^{33}$cm$^{-2}$s$^{-1}$
  - $E_{cm}$ = 6.5 TeV, L = 0.5·10$^{33}$cm$^{-2}$s$^{-1}$
- VLHC⊗LSC
  - $E_{cm}$ = 17 TeV, L > 10$^{33}$cm$^{-2}$s$^{-1}$
  - $E_{cm}$ = 24 TeV, L > 10$^{33}$cm$^{-2}$s$^{-1}$.

## 3. Linac-Ring Machines

The old idea [4] to collide a beam from a linear accelerator with a beam circulating in a storage ring has been recently renewed for two purposes:

1) to achieve the TeV scale at the constituent level in ep collisions [5-10]
2) to construct high luminosity particle factories [6, 11-13].

If future linear e$^+$e$^-$ colliders (or special e-linacs) are constructed near the existing (HERA, FNAL) or constructing (LHC) proton rings, a number of additional opportunities will arise. For example,

LHC⊗TESLA = LHC⊕TESLA
  ⊕ TeV scale ep Collider
  ⊕ TeV scale γp Collider
  ⊕ Multi-TeV scale e-nucleus Collider
  ⊕ Multi-TeV scale γ-nucleus Collider
  ⊕ FEL γ-nucleus (≡MeV energy laser in the nucleus rest frame).

### 3.1. Linac-Ring Type *ep* Colliders

It is known that synchrotron radiation restricts the electron energy obtainable at ring machines. A transition to linear accelerators seems unavoidable for $E_e$>100 GeV. For this reason HERA seems to be the first and last standard (ring-ring) type ep collider. The possible LHC×LEP has important disadvantage of $E_e/E_p<0.015$. Therefore, one should consider linac-ring type machines in order to achieve TeV scale at constituent level in lepton-hadron collisions. The possible alternative, namely µp colliders will face even more problems than the basic $\mu^+\mu^-$ collider.

Main parameters of linac-ring type ep colliders have been estimated in a number of papers [6-10, 14-16]. In Ref.[6], which deals with special e-linacs added to LHC and SSC, unrealistic parameters for proton beam ($\varepsilon_p$=0.02 mm×mrad, $n_p=10^{12}$ and $l_p$=10 cm) have been used. The rough estimations for UNK×VLEPP, LHC×CLIC and SSC×LSC are given in Ref.[7]. First serious paper on the subject is Ref.[8], where HERA×TESLA was considered in details, including interaction region layout.

### 3.1.1. Simplified Consideration

Luminosity of ep-collisions is given by

$$L_{ep} = \frac{n_p n_e}{s_{eff}} f_{eff}$$

where $n_p$ and $n_e$ are numbers of particles in corresponding bunches, $f_{eff}$ stands for collision frequency and $s_{eff}$ is effective transverse area at collision point.

There are two possible options for collision setup: in proton ring and on extracted proton beam [7]. The advantage of the first option is multiple usage of proton bunches, whereas in second option each proton bunch is used only once and can be maximally compressed. With recent design parameters of $e^+e^-$ colliders [2] the first option is preferable. In this case $f_{eff}=f_{rep}\times n_b$, where $f_{rep}$ is the repetition rate of electron pulses and $n_b$ is the number of bunches in a pulse. Of course, the bunch structure of electron and proton beams should be adjusted to each other.

In general, $s_{eff} = 4\pi\sigma_x^{eff}\sigma_y^{eff}$ and $\sigma_x^{eff}(\sigma_y^{eff})$ is biggest of $\sigma_x^e(\sigma_y^e)$ and $\sigma_x^p(\sigma_y^p)$. Usually $s_{eff} = s_p$, because electron bunches have much smaller transverse sizes, however one should be careful with $s_p$ vs $s_e^{min}$, obtained from beam-beam tune shift $\Delta Q_p$ [16]. Hereafter, let us restrict ourselves to the consideration of round beams. In this case

$$L_{ep} = \frac{n_p}{4\pi\varepsilon_p \beta_p^*} n_e f_{rep} n_b$$

where $\varepsilon_p$ is transverse emittance of proton beam and $\beta_p^*$ is amplitude function at interaction point.

Therefore, one should maximize $n_e f_{rep} n_b$ for electron beam and $n_p/\varepsilon_p\beta_p^*$ for proton beam, but

- $n_e f_{rep} n_b$ is constrained by electron beam power
- $n_p/\varepsilon_p\beta_p^*$ is constrained by
  - $\beta_p^* > l_p$ (see, however, Brinkmann-Dohlus ansatz [14])
  - emittance growth due to intrabeam scattering (IBS) [15].

At proton bunch length $l_p = 10$ cm, an acceptable value for number of protons in bunch is $n_p = 10^{11} \times \varepsilon_p^n/mm \times mrad$, where $\varepsilon_p^n = \gamma_p \times \varepsilon_p$ is normalized emittance ($\gamma_p - E_p/m_p$ is Lorentz factor). We need $n_p = 10^{12}$ at $\varepsilon_p^n = \pi \times mm \times mrad$, therefore *beam cooling in main ring* will be necessary in order to compensate $\varepsilon$ growth due to IBS.

### 3.1.2. Beam Separation

Interaction region layout depends on bunch spacing $\Delta t_s$. There are two options:
- head-on collisions for $\Delta t_s > 100$ ns [8]
- non-zero crossing angle for $\Delta t_s << 100$ ns.

First option corresponds to electron bunches from TESLA. In the second option crab-crossing will be needed.

In conclusion, modern accelerator technologies with reasonable future improvements will give opportunity to achieve TeV scale center-of-mass energies in ep-collisions at sufficiently high luminosities. For example, HERA×TESLA with $\sqrt{s} = 1$ TeV and $L_{ep} = 10^{31-32} cm^{-2}s^{-1}$ or LHC×TESLA with $\sqrt{s} = 6.5$ TeV and $L_{ep} = 10^{32-33} cm^{-2}s^{-1}$.

## 3.2. TeV Energy $\gamma p$ Colliders

The linac-ring type ep colliders are advantageous not only in $\sqrt{s}$ comparing with standard machines (HERA, LHC×LEP), but they also provide a unique possibility to construct $\gamma p$ colliders with practically the same $\sqrt{s}$ and luminosities [7, 22, 23].

### 3.2.1. High Energy $\gamma$-beam

Fifty years ago [17] Compton scattering by starlight quanta was investigated as a mechanism for the energy degradation of high-energy electrons in interstellar space. Twenty five years [18] later Compton backscattering of laser photons on extreme-relativistic electrons was proposed as a source of high energy photon beam. As the next stage the same mechanism was proposed in order to construct $\gamma e$ and $\gamma\gamma$ colliders [19] on the base of linear $e^+e^-$ machines (for present situation, see [20]). Finally, nonlinear effects in Compton scattering were observed using Nd:YAG laser beam on SLAC electron beam [21].

### 3.2.2. Main Parameters of $\gamma p$ Colliders

Below we follow in short the paper [23], where HERA×DLC, LHC×TESLA and LHC×e-linac based $\gamma p$ machines have been considered.

Why not standard type ep colliders?
- $E_{cm}$ limitations
- most important: $L_{\gamma p}/L_{ep} < 10^{-7}$, because each electron bunch is used only once.

Why not collisions on an extracted photon beam?
- because time spent of proton bunches is much smaller than filling time.

Therefore, one should consider the option with collisions in proton ring. In this case

$$L_{\gamma p} = 2 \frac{n_\gamma n_p}{s_p} f_{rep} n_b$$

where $n_\gamma = n_p$ (one to one conversion) and factor 2 reflects the fact that $s_\gamma \ll s_p$. As the result, we obtain $L_{\gamma p} = 2.5 \times 10^{31}$ cm$^{-2}$s$^{-1}$ and $\sqrt{s_{\gamma p}}^{max} = 1.16$ TeV for HERA×DLC, $L_{\gamma p} = 5 \times 10^{32}$ cm$^{-2}$s$^{-1}$ and $\sqrt{s_{\gamma p}}^{max} = 5.06$ TeV for LHC×TESLA and $\sqrt{s_{\gamma p}}^{max} = 2.27$ TeV for LHC×e-linac. These estimations do not take into account the effects of distance between the conversion region and the collision point, which can be itemized by followings:

- luminosity slowly decreases with increasing distance
- opposite helicity values for laser and electron beams are advantageous
- better monochromatization can be achieved by increasing the distance
- mean helicity of colliding photons approaches to one with increasing the distance

Recent results for HERA×TESLA and LHC×TESLA based γp colliders can be found in [24].

### 3.3. Multi-TeV Energy e-nucleus and γ-nucleus Colliders

It is known that LHC will operate also in nucleus-nucleus option [1]. The possibility of acceleration of different nucleus in HERA proton ring is investigated. Therefore, HERA×TESLA and LHC×TESLA will give opportunity to collide multi-hundred GeV energy electron and γ beams with multi-TeV energy nucleus beam. The main parameters of e-nucleus and γ-nucleus collisions were estimated in [25] (for recent situation, see [26]).

Within moderate improvements of nucleus beam parameters one may hope to obtain $L_{e-Pb} = 0.7 \cdot 10^{28}$ cm$^{-2}$s$^{-1}$ and $L_{\gamma Pb} = 1.3 \cdot 10^{28}$ cm$^{-2}$s$^{-1}$ at LHC×TESLA. These values correspond to $L_{e-N} = 1.3 \cdot 10^{30}$ cm$^{-2}$s$^{-1}$ and $L_{\gamma N} = 2.6 \cdot 10^{30}$ cm$^{-2}$s$^{-1}$ at nucleon level. It is possible that more radical improvements will increase these numbers by an order.

### 3.4. FEL γ-nucleus Colliders

The TESLA can operate as Free Electron Laser in X-ray region. Colliding of FEL beam with nucleus bunches may give a unique possibility to investigate "old" nuclear phenomena in rather unusual conditions. Indeed, keV energy FEL photons will be seen in the rest frame of nucleus as the MeV energy "laser" beam. Moreover, since the accelerated nucleus is fully ionized, we will be free from possible background induced by low-shell electrons. This option needs more investigations from both accelerator and nuclear physics viewpoints.

## 4. Some Speculations on TeV Scale Physics

Physics at TeV scale is the subject of large number of scientific papers and reviews (see, for example, CERN Yellow Reports and Snowmass Proceedings). Below I restrict myself to four items, which have not received wide recognition, but are important for future TeV energy colliders. Short remarks on each item are presented and for details I refer to [27] and original papers sited below.

### 4.1. The Fourth SM Family

Twenty years ago the *flavor democracy* was proposed [28] in order to solve some problems of the Standard Model. However, in the three SM family case this approach leads to a number of unacceptable results, such as a low value of t-quark mass etc. On the other hand, flavor democracy seems very natural in the framework af SM and problems disappear if the fourth fermion family is introduced [29-31].

Let us present the main assumptions (at this stage we assume that neutrinos are Dirac particles):

1. Before the spontaneous symmetry breaking fermions with the same quantum numbers are indistinguishable. Therefore, Yukawa couplings are equal within each type of fermions ($a_{ij}^d \equiv a^d$, $a_{ij}^u \equiv a^u$, $a_{ij}^l \equiv a^l$ and $a_{ij}^\nu \equiv a^\nu$, where $i$ and $j$ denote family number) and in the SM basis one deals with four 4×4 mass matrices all elements of which are equal.
2. There is only one Higgs doublet, which gives Dirac masses to all four types of fermions. Therefore, Yukawa constants for different types of fermions should be (nearly) equal ($a^d \approx a^u \approx a^l \approx a^\nu \approx a$).
3. $a$ lies between $\sqrt{(4\pi\alpha_{em})}$ and $g_w$ (with preferable value $a \approx g_w$).

As the result, fourth family fermions receive the masses $m_4 \approx 4g_w\eta = 8m_W \approx 640$ GeV, while first three families are massless. In order to provide masses for quarks and charged leptons from the first three families minimal deviations from full democracy was considered in [31], where CKM matrix elements have been calculated using quark masses as input parameters. Results are in good agreement with experimental data.

The main decay modes of the fourth family fermions are predicted to be $u_4 \to b+W^+$, $d_4 \to t+W^-$, $l_4 \to \nu_\tau+W^-$ and $\nu_4 \to \tau+W^+$. The fourth family quarks will be copiously produced at LHC [32], whereas the best places to search for the fourth family leptons are future linear $e^+e^-$ machines (including $\gamma\gamma$ option for fourth charged lepton search) and $\mu^+\mu^-$ collider.

The existence of fifth SM family seems unnatural because of large value of t-quark mass and LEP results on neutrino counting, which showed that there are only three "light" non-sterile neutrinos, whereas five family SM predicts four "light" Dirac neutrinos.

## 4.2. Compositness vs SUSY or Compositness & SUSY

It is known that the number of free observable parameters put by hand in SN is equal to 26 in the three family case and 40 in the four family case (DMM approach reduces these numbers to 20 and 28, respectively), if neutrinos are Dirac particles. The natural question is: "How many free parameters contains minimal supersimmetric extension of the Standard Model (MSSM)?"

### 4.2.1. CKM Mixings in MSSM

The numbers of observable mixing angles and phases in n family SM are given by well-known formulae:

$$N_\theta = \frac{n(n-1)}{2}, \quad N_\varphi = \frac{(n-1)(n-2)}{2}.$$

Let us estimate corresponding numbers in MSSM [33]. In the framework of $SU_C(3)\times SU_W(2)\times U_Y(1)$ model with $l$ up quarks and $m$ down quarks, whose left-handed components form $n$ weak isodoublets, we obtain [34] (following Kobayashi-Maskawa arguments [35]):

$$N_\theta = \frac{n(n-1)}{2} + n(l+m-2n), \quad N_\varphi = \frac{(n-1)(n-2)}{2} + (n-1)(l+m-2n).$$

For the $n$-generation $E_6$-induced model ($m = 2l = 2n$) we obtain $N_\theta = n(3n-1)/2$ and $N_\varphi = (n-1)(3n-2)/2$. In three family case this gives 12 mixing angles and 7 phases in quark sector.

By following similar arguments for $n$ family MSSM one can obtain

$$N_\theta^{q,\tilde{q}} = N_\varphi^{q,\tilde{q}} = n(5n-3).$$

If neutrinos are Dirac particles, the same number of free parameters

$$N_\theta^{l,\tilde{l}} = N_\varphi^{l,\tilde{l}} = n(5n-3)$$

comes from lepton-slepton sector. Moreover, we have also 12$n$ mass values for leptons, sleptons, quarks and squarks. In addition, one has:
- 2 angles and 4 phases from chargino diagonalization
- 6 angles and 10 phases from neutralino diagonalization
- …

Total number of free parameters is $N > 20n^2+22$, i.e. N > 202 for three family and N > 342 for four family MSSM. Let me remind that number of free parameters in three family SM without right-handed neutrinos being 19 was one of the main arguments to go Beyond the Standard Model!

Message: *SUSY should be realized at a more fundamental level.*

Today, there are two favorite candidates:

1. Preonic level!
2. SUGRA?

### 4.2.2. Supersymmetric Preonic Models of Quarks and Leptons

There are at least two arguments favoring compositeness:

1. SUSY GIM cancellation ($K_L$-$K_S$ transition etc.) requires $\delta m_{\tilde{q}}^2 \approx \delta m_q^2$ ($m_{\tilde{u}}^2 - m_{\tilde{c}}^2 \approx m_c^2$ etc.) and $U_{CKM}^{\tilde{q}} \approx U_{CKM}^{q}$. This seems natural in preonic models.
2. MSSM includes two observable phases even in the simplest case of one family: $N_\theta = N_\varphi = 2$ for $n = 1$.

Composite models of leptons and quarks can be divided into two classes: fermion-scalar models and three-fermion models. Let us briefly consider main consequences of SUSY extensions for these classes. Below we present the simplified options where only one superpartner for each preon is introduced and flavour mixings are absent (according to N=1 SUSY each charged fermion has two superparetners etc.). More realistic versions will be considered in details elsewhere [36].

In the first class, SM fermions (quarks and leptons) are composites of scalar preons, denoted by $S$, and fermion preons, denoted by $F$. In minimal variant $q, l = \{FS\}$. In principle, there are two opportunities:

- scalar preons are superpartners of fermion preons
- each preon has its own superpartner.

The second option leads to the quadrupling of SM matter fields (instead of doubling in MSSM). One has following states: SM fermion ($FS$) with $m \sim 0$, scalar ($\tilde{F}S$) with $m \sim \mu$, scalar ($F\tilde{S}$) with $m \sim \mu$ and fermion ($\tilde{F}\tilde{S}$) with $m \sim 2\mu$.

In the second class, quarks and leptons are composites of three fermionic preons and each of them has at least seven partners. In other words we have: SM fermion ($F_1 F_2 F_3$) with $m \sim 0$; three scalars ($\tilde{F}_1 F_2 F_3$), ($F_1 \tilde{F}_2 F_3$) and ($F_1 F_2 \tilde{F}_3$) with $m \sim \mu$; three fermions ($\tilde{F}_1 \tilde{F}_2 F_3$), ($\tilde{F}_1 F_2 \tilde{F}_3$) and ($F_1 \tilde{F}_2 \tilde{F}_3$) with $m \sim 2\mu$; and scalar ($\tilde{F}_1 \tilde{F}_2 \tilde{F}_3$) with $m \sim 3\mu$.

Of course, mixings between quarks (leptons, squarks, sleptons) can (and should?!) drastically change the simple mass relations given above. Therefore, it is quite possible that the search for SUSY at LHC will give rather surprising results.

### 4.2.3. General Remarks on Composite Models

In principle, one can consider four stages of compositness (each stage includes previous ones):

i) Composite Higgs
ii) Composite quarks and leptons
iii) Composite W- and Z- bosons
iv) Composite photon and gluons (?)

**New particles:** The well-known representative of the first stage is Technicolor Model, which gives masses to W- and Z- bosons in a best manner but has serious problems with fundamental fermion masses (Extended Technicolor etc.). Therefore, one should deal at least with the second stage. In this case model predicts a number of new particles with rather unusual quantum numbers: excited quarks and leptons, leptoquarks (HERA events?), colour-sextet quarks and color-octet leptons. If the third stage is realized in Nature, excited W and Z bosons, colour octet W and Z, scalar W ans Z are predicted also. The realization of the fourth stage seems today less natural because photon and gluons correspond to the unbroken gauge symmetries.

The masses of new particles are expected to lie in the range of compositness scale $\Lambda$, which exceeds TeV. Of course, if SUSY takes place at preonic level all these new particles have a number of (SUSY) partners.

Finally, it is quite possible that SUSY is realized at pre-preonic level!

**New interactions:** Nobody knows real dynamics, which keeps preons together to form SM particles. Today, the most popular candidate is hypercolor (some extension of QCD). However, it is quite possible that new dynamics is based on concepts, which differ drastically from the known ones (like the difference between quantum and classic physics). In any case, we expect that some residual "contact" (Fermi-like) interactions should manifest themselves at scale smaller than $\Lambda$ with intensity proportional to $1/\Lambda^2$.

### 4.3. SUGRA Manifestations

As we mentioned in previous subsection, the second favorite candidate to solve problems unsolved by Standard Model is SUGRA, which simultaneously unifies al known fundamental interactions including gravitation. Unfortunately, SUGRA does not solve the masses and mixings problems (at least for today).

The most realistic scenarios from SUGRA scale to SM scale predict the existence of at least one additional neutral intermediate vector boson with mass in the region 1 to 10 TeV (see, for example, [37] and references therein). The discovery limit for new $Z^{'}$ boson at LHC is about 5 TeV. The search for indirect manifestations of $Z^{'}$ at future lepton colliders and linac-ring type ep colliders will be sensitive up to 20 TeV.

#### 4.3.1. Isosinglet Quarks

The first family fermion sector of the $E_6$-induced model has the following $SU_C(3) \times SU_W(2) \times U_Y(1)$ structure:

$$\begin{pmatrix} u_L \\ d_L \end{pmatrix} \quad u_R \quad d_R \quad D_L \quad D_R \quad \begin{pmatrix} \nu_{eL} \\ e_L \end{pmatrix} \quad \nu_{eR} \quad e_R \quad \begin{pmatrix} N_{eL} \\ E_L \end{pmatrix} \quad \begin{pmatrix} N_{eR} \\ E_R \end{pmatrix} \quad N_e$$

New isosinglet quarks decay only due to mixings with usual down-type quarks d, s and b. Remember that quark sector in three family case contains 12 observable mixing angles and 7 observable phases. Let us suppose that interfamily mixing is dominant and usual CKM mixings lie in up-quark sector. Then, in weak base one has

$$\begin{pmatrix} u_L^{CKM} \\ d_L^{\varphi} \end{pmatrix} \quad u_R \quad d_R \quad D_L^{\varphi} \quad D_R$$

where

$$\begin{cases} d^{\varphi} = d\cos\varphi + D\sin\varphi \\ D^{\varphi} = -d\sin\varphi + D\cos\varphi \end{cases}$$

and $\sin^2\varphi \ll 1$. For $m_D - m_W \approx m_D - m_Z \gg m_Z - m_W$ one has BR(D $\to$ u + W) $\approx$ 0.6 and BR(D $\to$ d + Z) $\approx$ 0.4. Therefore, we expect BR(D $\to$ jet + $l^+l^-$) $\approx$ 0.012 and BR(D $\to$ jet + $\nu\nu$) $\approx$ 0.072 for decay modes which differ isosinglet quarks from the fourth SM family quarks. LHC with $L^{int} \approx 10^5 pb^{-1}$ will produce $\approx 5\cdot 10^5 (3\cdot 10^4)$ DD pairs per year if $m_D$=0.5(1.0) TeV.

Decays of new charged and neutral leptons strongly depend on their mass pattern. In general, leptonic sector also contains flavor changing neutral currents. If new Z' boson is sufficiently light ($m_{Z'} < \sqrt{s}$) and (some of the) new leptons and quarks have masses less than $m_{Z'}/2$, future lepton colliders will give a unique opportunity for investigation of their properties.

### 4.3.2. Flavour Democracy

In the case of three $E_6$ families, quark sector of the model has the form

$$\begin{pmatrix} u_L \\ d_L \end{pmatrix} \quad u_R \quad d_R \quad D_L \quad D_R \quad \begin{pmatrix} c_L \\ s_L \end{pmatrix} \quad c_R \quad s_R \quad S_L \quad S_R \quad \begin{pmatrix} t_L \\ b_L \end{pmatrix} \quad t_R \quad b_R \quad B_L \quad B_R$$

According to flavor democracy we deal with following mass matrices for up and down quarks

$$M_u = a_u \eta \begin{pmatrix} 1 & 1 & 1 \\ 1 & 1 & 1 \\ 1 & 1 & 1 \end{pmatrix}$$

and

$$M_d = a_d \eta \begin{pmatrix} 1 & 1 & 1 & 1 & 1 & 1 \\ 1 & 1 & 1 & 1 & 1 & 1 \\ 1 & 1 & 1 & 1 & 1 & 1 \\ k & k & k & k & k & k \\ k & k & k & k & k & k \\ k & k & k & k & k & k \end{pmatrix}$$

where $k = \mu / a_d \eta$ and $\mu$ is the next to SM scale ($\mu \gg 100$ GeV). As a result we obtain: $m_t = 3g_u\eta$ and $m_u = m_c = 0$ in up sector, $m_B = 3g_d\eta + 3\mu$ and $m_d = m_s = m_b = m_D = m_S = 0$ in down sector. After breaking of flavor democracy it is natural to expect that $m_D \ll m_S \ll \mu$. Therefore, with high probability at least one isosinglet quark (D-quark) will be covered by LHC.

For the similar reasons, at least one new charged lepton (E$^-$) will be covered by future lepton colliders.

### 4.4. An Example of "Unexpected" New Physics

Let me conclude this section with a short remark on the consequences of possible low-energy compactification, that is new (space) dimensions at TeV (??) scale. The work on the subject is under development [38] and presented results are (very?) preliminary. Firstly, the existence of two new dimensions seems preferable, in other case one faces problems in formulation of SM. "Infinite" nimber of "electrons", "muons" and other fundamental particles are predicted. The mass spectrum depends on compactification mechanism and there are two extreme cases: $m_{e(n)} = n \times M$ and $m_{e(n)} = n^2 \times M$, where $M \sim 1/r$ denotes the compactification scale. The usual electron corresponds to $e_0$. Same relations take place for other SM particles. In principle, new particles like "heavy electrons" are expected to be stable and should be produced pairly, but…

### 5. Physics at TeV Energy *ep* and *γp* Colliders

Although physics search programs of new ep and γp colliders are much less developed than those of LHC, NLC and $\mu^+\mu^-$ colliders, a lot of papers on this subject were published during the last decade. The physics at UNK×VLEPP based ep and γp colliders were considered in [5] and [39, 40], correspondingly. Resonant production of excited quarks at γp colliders was investigated in [41]. Reference [42] dealt with physics at future γγ, γe and γp colliders. Wino production at HERA×LC based γp collider was considered in[43]. Today, the main activity on this subject is concentrated in Ankara University HEP group [44-55]. In Reference [56] we review physics search potential of HERA×LC based γp collider. Recently, Higgs boson production at LHC×LEP based (?) γp collider has been studied in [57] and [58], however their results should be recalculated because (as argued in [23]) γp colliders can be constructed only on the base of linac-ring type ep machines.

## 5.1. Physics at Linac-Ring Type *ep* Colliders

This topic was sufficiently developed during preparation of HERA and study of LHC×LEP physics search potential. Linac-ring type machines will give opportunity to investigate appropriate phenomena at

- higher center of mass energies,
- better kinematic conditions.

The situation is illustrated in Table 1.

**Table 1.** Center of mass energies and kinematics of ep colliders

| Machine | $\sqrt{s_{ep}}$, TeV | $E_e/E_p$ |
|---|---|---|
| HERA | 0.3 | 1/30 |
| LHC×LEP | 1.2 | 1/120 |
| HERA×TESLA | 1(2.4) | 1/4 |
| LHC×TESLA | 2.6(6.5) | 1/5 |
| VLHC×LSC | 17(24) | 1/6 |

Let us remind that confirmation of recent results [59] from HERA will favor new ep machines. Physics search program of HERA×TESLA based ep collider is considered in [60].

## 5.2. Physics at $\gamma p$ Colliders

Below we illustrate physics search potential of future γp machines. As samples we use HERA×TESLA (1 TeV×0.3 TeV) with $L_{\gamma p}^{int}$ = 500 pb$^{-1}$ and LHC×TESLA (7 TeV×1.5 TeV) with $L_{\gamma p}^{int}$ = 5 fb$^{-1}$.

### 5.2.1. SM Physics

- Total cross-section at TeV scale can be extrapolated from existing low energy data as $\sigma(\gamma p \rightarrow hadrons) \sim 100 \div 200 \mu b$, which corresponds to ~10$^{11}$ hadronic events per working year

- Two-jet events (large $p_t$)
    HERA×TESLA: 10$^4$ events with $p_t$ > 100 GeV
    LHC×TESLA: 10$^4$ events with $p_t$ > 500 GeV

- $t\bar{t}$ pair production
    HERA×TESLA: 10$^3$ events per year
    LHC×TESLA: 10$^5$ events per year

- $\bar{b}b\,(\bar{c}c)$ pair production
    HERA×TESLA: $10^8$ events
    LHC×TESLA: $10^9$ events
  The region of extremely small $x_g \sim 10^{-6} \div 10^{-7}$ can be investigated (phenomenon of inverse evolution of parton distributions)

- W production
    HERA×TESLA: $10^5$ events
    LHC×TESLA: $10^6$ events
  $\Delta\kappa_W$ can be measured with accuracy of 0.01 (0.001 taking into account $\gamma$ polarization?)

- Higgs boson production ($\gamma p \rightarrow WH + X$)
    HERA×TESLA: 20 events at $m_H = 100$ GeV
    LHC×TESLA: 1000 events at $m_H = 100$ GeV and 100 events at $m_H = 300$ GeV

- Fourth SM family quarks (discovery limits for 100 events per year)
    HERA×TESLA: $m_{u4} = 250$ GeV, $m_{d4} = 200$ GeV
    LHC×TESLA: $m_{u4} = 1000$ GeV, $m_{d4} = 800$ GeV

### 5.2.2. Beyond the SM Physics

Below we present discovery limits for 100 events per year:

- $\gamma p$ colliders are ideal machines for $u^*$, $d^*$ and $Z_8$ search
    HERA×TESLA: $m_{u*} = 0.9$ TeV, $m_{d*} = 0.7$ TeV, $m_{Z8} = 0.7$ TeV
    LHC×TESLA: $m_{u*} = 5$ TeV, $m_{d*} = 4$ TeV, $m_{Z8} = 4$ TeV

- single $l_q$ production
    HERA×TESLA: 0.7 TeV
    LHC×TESLA: 3 TeV

- pair $l_q$ production
    HERA×TESLA: 0.3 TeV
    LHC×TESLA: 1.7 TeV

- SUSY should be realized at preonic level, however for MSSM particles we have (neglecting SUSY CKM mixings)

|  |  | HERA×TESLA | LHC×TESLA |
|---|---|---|---|
| $\gamma p \to \tilde{W}\tilde{q} + X$ | $m_{\tilde{W}} = m_{\tilde{d}}$ | 0.25 TeV | 0.9 TeV |
|  | $m_{\tilde{d}}$ ($m_{\tilde{W}} = 0.1$ TeV) | 0.5 TeV | 2 TeV |
|  | $m_{\tilde{W}}$ ($m_{\tilde{d}} = 0.1$ TeV) | 0.3 TeV | 1.2 TeV |
| $\gamma p \to \tilde{g}\tilde{q} + X$ | $m_{\tilde{g}} = m_{\tilde{d}}$ | 0.2 TeV | 0.8 TeV |
|  | $m_{\tilde{g}}$ ($m_{\tilde{q}} = 0.1$ TeV) | 0.4 TeV | 2 TeV |
|  | $m_{\tilde{q}}$ ($m_{\tilde{g}} = 0.1$ TeV) | 0.3 TeV | 1 TeV |
| $\gamma p \to \tilde{\gamma}(\tilde{Z})\tilde{q} + X$ | $m_{\tilde{\gamma}} = m_{\tilde{q}}$ | 0.15 TeV | 0.2 TeV |
|  | $m_{\tilde{q}}$ ($m_{\tilde{\gamma}} = 0.1$ TeV) | 0.2 TeV | 0.4 TeV |
|  | $m_{\tilde{q}}$ ($m_{\tilde{Z}} = 0.1$ TeV) | 0.17 TeV | 0.3 TeV |
| $\gamma p \to \tilde{q}^c \tilde{q} + X$ |  | 0.25 TeV | 0.8 TeV |

**5.3. Physics at γ-nucleus Colliders**

Center of mass energy of LHC×TESLA based γ-nucleus collider corresponds to $E_\gamma \sim$ PeV in the lab system. At this energy range cosmic ray experiments have a few events per year, whereas γ-nucleus collider will give few billions events. Very preliminary list of physics goals contains:

- total cross-sections to clarify real mechanism of very high energy γ-nucleus interactions
- investigation of hadronic structure of photon in nuclear medium
- according to VMD, proposed machine will be also ρ-nucleus collider
- formation of the quark-gluon plasma at very high temperatures but relatively low nuclear density
- gluon distribution at extremely small $x_g$ in nuclear medium ($\gamma A \to Q\bar{Q} + X$)
- investigation of both heavy quark and nuclear medium properties ($\gamma A \to J/\Psi(Y) + X$, $J/\Psi(Y) \to l^+l^-$)
- existence of multi-quark clusters in nuclear medium and few-nucleon correlations.

## 6. Conclusion or Dreams for Next Century

There are strong arguments favoring that the rich spectrum of new particles and/or interactions will manifest themselves at TeV scale. An exploration of this scale at constituent level will require all possible types of colliding beams. Today, work on physics search programs and machine parameters for future hadron and lepton colliders is quite advanced, whereas those for linac-ring type lepton-hadron colliders need an additional R&D. For this reason we suggest to organize two workshops on lepton-hadron machines – one on physics goals and other on machine parameters – in the next year. Then, an International Conference on "TeV Scale: Physics and Machines" will be very useful for long-term planning in the field of High Energy Physics. In this context, one should compare physics at:

- Upgraded Tevatron (2 TeV), NLC (0.5 TeV) and $\mu^+\mu^-$ (0.5 TeV) with HERA×TESLA (1÷2 TeV) ep and $\gamma$p colliders
- LHC (14 TeV) and NLC (1.5 TeV) with LHC×TESLA (4÷6 TeV)
- VLHC (60 TeV), LSC (5 TeV) and $\mu^+\mu^-$ (4 TeV) with VLHC×LSC (17÷24 TeV) and $\mu$p (15 TeV).

Let me conclude with the following tables, which reflect (personal) dreams for early 21$^{st}$ century:

**Table 2.** Near Future (2010)

|  | Colliding beams | $\sqrt{s}$, TeV | L, $10^{32}$cm$^{-2}$s$^{-1}$ | $\sqrt{\hat{s}}$, TeV |
|---|---|---|---|---|
| LHC | pp | 14 | 100 | 3÷4 |
| NLC1 | $e^+e^-$($\gamma$e, $\gamma\gamma$) | 0.5(0.4) | 10 | 0.5(0.4) |
| NLC2 | $e^+e^-$($\gamma$e, $\gamma\gamma$) | 1.5(1.2) | 100 | 1.5(1.2) |
| $\mu^+\mu^-$ | $\mu^+\mu^-$ | 0.5 | 5 | 0.5 |
| HERA×TESLA | ep($\gamma$p) | 1(0.9) | 1 | ~0.6(0.5) |
| LHC×TESLA | ep($\gamma$p) | 5(4) | 5 | ~3(2) |

**Table 3.** World Laboratory (2020)

| Colliding beams | pp | ee ($\gamma$e, $\gamma\gamma$) | $\mu^+\mu^-$ | ep | $\gamma$p | $\mu$p |
|---|---|---|---|---|---|---|
| $\sqrt{s}$, TeV | 60 | 5 (4, 4.5) | 4 | 17(24) | 15(22) | 15 |
| L, $10^{32}$cm$^{-2}$s$^{-1}$ | 100 | 1000 | 1000 | 100 | 100 | 100 |
| $\sqrt{\hat{s}}$, TeV | ~10 | 5 (4, 4.5) | 4 | ~8 | ~7 | ~7 |

**Acknowledgements**

The main part of this talk was prepared and presented at DESY seminar last summer. I am grateful to B.H. Wiik for invitation, support and stimulating discussions; to D. Trines

# PART II. Ankara Workshop on Linac-Ring Type *ep* and $\gamma p$ Colliders

### 7.1. Preface

The first International Workshop on Linac-Ring Type *ep* and $\gamma p$ Colliders held in Ankara between 9-11 April 1997. The workshop was organized with the supports from Scientific and Technical Research Council of Turkey (TUBITAK), Ankara University and Deutsches Elektronen-Synchrotron DESY Directorate. During the workshop more than thirty reports have been presented and most of the are published in this proceedings. The recently proposed lepton-hadron machines, which open a fourth way to investigate TeV scale physics, are discussed thoroughly from the machine and physics aspects.

**Editors:**
S. Atag (Ankara University)
S. Turkoz (Ankara University)
A.U. Yilmazer (Ankara University)

### 7.2. International Advisory Committee
M. Atac (FNAL)
S. Ayik (Tennessee University)
E. Boos (Moscow State University)
S. Gershtein (IHEP, Protvino)
L. Okun (ITEP, Moscow)
N.K. Pak (TUBITAK and METU, Ankara)
V. Savrin (Moscow State University)
S. Sultansoy (Ankara University and Azerbaijan Academy of Sciences)
D. Trines (DESY)
A. Wagner (DESY)
B.H. Wiik (DESY)
C. Yalcin (METU, Ankara)

### 7.3. Organizing Committee
S. Atag (Ankara University)
Z.Z. Aydin (Ankara University, Chairman)
A. Celikel (Ankara University)
A.K. Ciftci (Ankara University)
O. Yavas (Ankara University)

### 7.4. Workshop Conclusion and Recommendations

New linac-ring type *ep*, $\gamma p$ and $\mu p$ colliders will be constructed after operation of basic $e^+e^-$, $\gamma e$, $\gamma\gamma$, *pp* and $\mu^+\mu^-$ colliders. They have advantages in study of quarks and gluons in the proton because probing particles (*e*, $\gamma$, $\mu$) have in general well known structures. These type of collisions are also optimum for production and study of some new particles such as leptoquarks. The expexted luminosity of *ep* and $\gamma p$ colliders is lower than that of above mentioned basic colliders. Nevertheless, there are a number of physical problems, which can be solved at these new type colliders. These are

- QCD in the new region of parameters

- Leptoquarks, leptogluons and new contact interactions
- Searching for SUSY and wide spectrum of problems beyond the SM, etc.

In order to obtain here really new results complementary to those at basic colliders, the luminosities $L(ep) \geq 10^{31}$ and $L(\gamma p) \geq 10^{30}$ are necessary in units of cm$^{-2}$s$^{-1}$ and seem to be sufficient. Higher luminosities require cooling of the proton beams which needs additional studies. Concerning the $\mu p$ colliders rough estimates give the luminosity $L(\mu p) = 10^{33}$ cm$^{-2}$s$^{-1}$, however this topic calls for more detailed investigation.

As a result of the workshop, *participants came to the point that it will be useful to organize two workshops, one on the machine parameters and the other on the physics research program, during the next year.*

# PART III. Appeal to ICFA, ECFA and ACFA

Today, different aspects of the TESLA*HERA based ep and γp colliders are investigated by an informal THERA group. However, due to comparatively low values of center of mass energy (~1 TeV) and luminosity ($10^{30\div31}$cm$^{-2}$s$^{-1}$), THERA will be effective mainly in investigation of small $x$ region. Of course, this research is very important for our understanding of QCD (in some sense the investigation of small $x$ region for strong interactions is analog of the Higgs search for electroweak interactions). Nevertheless, concerning high energy linac-ring type machines, the TESLA*LHC complex seems to be the most promising one for middle-term future.

**Dear Colleagues,** *may I ask you to organize the Common ICFA-ECFA-ACFA Study Group on Linac-Ring Type ep, γp, eA, γA and FEL γA Colliders*?

**References (after 1997)**

1. S. Sultansoy, *The post-HERA era: Brief Review of Future Lepton-Hadron and Photon-Hadron Colliders,* DESY preprint DESY-99-159 (Oct 1999). [hep-ph 9911417]

2. THERA Homepage: www.ifh.de/thera

3. O. Yavas, A.K. Ciftci and S. Sultansoy, *TESLA*HERA as Lepton (Photon) – Hadron Collider,* Ankara University preprint AU-HEP-00-03 (Apr 2000), Contributed paper to EPAC2000. [hep-ex 0004013]

4. A.K. Ciftci, S. Sultansoy and O. Yavas, *Linac*LHC based ep, γp, eA, γA and FEL γA Colliders: Luminosity and Physics.* [hep-ex 0006030]

5. A.K. Ciftci, S. Sultansoy and O. Yavas, *TESLA*HERA based γp and γA Colliders,* Presented by S. Sultansoy at the Int. Workshop on High Energy Photon Colliders (14-17 June 2000, DESY, Hamburg, Germany). [hep-ex 0007009]